\documentclass[sn-nature]{sn-jnl}
\usepackage{xr} 
\usepackage{tabularx}
\usepackage{csquotes}
\usepackage{rotating}
\usepackage{epigraph}
\usepackage{float}
\usepackage[dvipsnames]{xcolor}
\usepackage[greek,english]{babel}
\usepackage{graphicx}%
\usepackage{multirow}%
\usepackage{amsmath,amssymb,amsfonts}%
\usepackage{amsthm}%
\usepackage{amsmath}
\usepackage{amssymb}
\usepackage{mathrsfs}%
\usepackage[title]{appendix}%
\usepackage{xcolor}%
\usepackage{textcomp}%
\usepackage{manyfoot}%
\usepackage{booktabs}%
\usepackage{lmodern}%
\usepackage{algorithm}%
\usepackage{algorithmicx}%
\usepackage{algpseudocode}%
\usepackage{listings}%

\theoremstyle{thmstyleone}%

\begin{document}

\title[Article Title]{Cellular Development Follows the Path of Minimum Action}


\author*[1]{\fnm{Rohola} \sur{Zandie}}\email{rohola@mit.edu}

\author[1]{\fnm{Farhan} \sur{Khodaee}}

\author[1]{\fnm{Yufan} \sur{Xia}}

\author[1,2]{\fnm{Elazer} \sur{R. Edelman}}

\affil*[1]{\orgdiv{Institute for Medical Engineering and Science}, \orgname{Massachusetts Institute of Technology}, \orgaddress{\city{Cambridge}, \postcode{02139}, \state{MA}, \country{USA}}}

\affil[2]{\orgdiv{Department of Medicine (Cardiovascular Medicine)}, \orgname{Brigham and Women’s Hospital}, \orgaddress{\city{Boston}, \postcode{02115}, \state{MA}, \country{USA}}}

\abstract{Cellular development follows a stochastic yet rule-governed trajectory, though the underlying principles remain elusive. Here, we propose that cellular development follows paths of least action, aligning with foundational physical laws that govern dynamic systems across nature. We introduce a computational framework that takes advantage of the deep connection between the principle of least action and maximum entropy to model developmental processes using Transformers architecture. This approach enables precise quantification of entropy production, information flow curvature, and local irreversibility for developmental asymmetry in single-cell RNA sequence data. Within this unified framework, we provide interpretable metrics: entropy to capture exploration-exploitation trade-offs, curvature to assess plasticity–elasticity dynamics, and entropy production to characterize dedifferentiation and transdifferentiation. We validate our method across both single-cell and embryonic development datasets, demonstrating its ability to reveal hidden thermodynamic and informational constraints shaping cellular fate decisions.
}





\maketitle

\section{Introduction}
\label{sec:introduction}

Cellular development is a complex biological process critical for a wide range of physiological functions, ranging from embryonic growth and cellular differentiation to tissue repair and regeneration. Over the centuries, many researchers have sought a universal principle explaining how cells make developmental \textit{decisions}. Karl Ernst von Baer was the first to systematically study cellular development in this light, and in 1828 he introduced empirical evidence for what became known as the ``Law of Development'' \cite{von1828entwickelungsgeschichte,abzhanov2013baer}. Building on von Baer’s foundational work, Wilhelm Roux (1850–1924) and Hans Driesch (1867–1941) ushered in the era of experimental embryology, further probing the rules of morphogenesis \cite{maienschein1991origins}. It was, however, Conrad Waddington who bridged experimental observations with genetics by proposing the ``epigenetic landscape'' as the conceptual field in which cellular development unfolds \cite{waddington2014strategy, waddington2015animals}. Although conceptually powerful, this qualitative framework does not yet offer the quantitative metrics and mechanistic details needed to fully comprehend cellular decision-making \cite{shi2020quantifying}.


A clearer path to understanding the time evolution of biological development may lie in differentiating between the notion of a \textit{process} and that of an \textit{object}. In contrast to approaches in physics, which often focus on tracking a single \textit{object} over time, biological systems typically involve collections of entities evolving through shared, more generic \textit{processes} \cite{bailly20112}. Thus organisms are specific \textit{objects} resulting from more universal developmental \textit{processes}. Likewise, at the cellular level, individual cells, each with its own developmental path, known as cell fate, can be understood as \textit{particular realizations} of the same overarching process. Focusing on this process-oriented perspective is a crucial step toward establishing more rigorous and quantitative principles for how cells make developmental decisions. 

In this paper, we propose a more general mathematical representation of these developmental processes using \textbf{random process} $\left\{X_t\right\}_{t \in T}$, where $t$ indexes time. Each realization of this process defines a \emph{trajectory} $\Gamma \equiv\left\{X_t: t \geq 0\right\}$, which, in principle, is a continuous functions. The collection of all such trajectories constitutes the experimental observations of the system’s dynamics. However, in practice, we do not observe an infinite set of trajectories or continuous-time data; instead, we work with a finite number of trajectories, each sampled at discrete time points. Specifically, we rely on \textbf{finite samples} obtained from a parametric joint probability distribution $p_\theta(X_0, \cdots, X_N)$ over a chosen time window of size $N$. Formally, each specific trajectory $\gamma$ is drawn from this distribution:

\begin{equation}
    \gamma=(x_0, \cdots x_N) \sim p_\theta(X_0, \cdots, X_N)
\end{equation}

It is important to note that this modeling approach treats each trajectory as an indivisible building block rather than assuming a \textit{memoryless Markov chain}. This distinction is crucial for our later modeling, which prioritizes the \textbf{primacy of trajectories} in the analysis rather than individual time points. By focusing on entire trajectories, we capture the inherent dependencies and long-range correlations in the system, which are essential for accurately representing biological processes \cite{stumpf2017stem}.

To provide additional insight into how trajectories might be selected or weighted, we use \emph{Lagrangian mechanics} in physics which offers a powerful framework for finding a system’s ``right'' trajectory by minimizing an action functional. In classical physics, this yields a unique path dictated by the laws of motion, whereas in quantum physics, Feynman’s \textbf{path integral formulation} generalizes this principle by integrating over infinitely many possible trajectories between two time points. 

This framework is not limited to these physical systems but can be viewed as a universal principle applicable to complex systems operating under probabilistic constraints, providing a natural foundation for modeling stochastic biological processes \cite{kaila2008natural}.

Unlike physics, explicit formulation of biological paths has not been possible. Therefore, we focus on data-driven computational modeling. We first start with the Lagrangian formulation applied to biological processes and develop a powerful computational framework that is general yet interpretable enough to explain different developments such as cell differentiation and reprogramming \cite{schiebinger2019optimal}, spatiotempral mouse organogenesis development \cite{chen2022spatiotemporal} and mouse embryonic development \cite{qiu2024single}, from gastrula to pup using single cell RNA-seq data. This can be achieved by bridging the gap between the \textbf{principle of least action} and the \textbf{principle of maximum entropy} (MaxEnt) using the computational framework of autoregressive neural networks. In addition, this framework helps us to identify cellular development as ``geodesics'' on the configuration manifold induced by the interactions in the gene expression space. We demonstrate how the geometry of this manifold, as encoded in its curvatures, reveals crucial insights into the flow of information throughout the process. Our results demonstrate a quantitative understanding of development and concepts such as irreversibility, entropy, and the curvature of development processes with minimum assumptions.

\begin{figure}
    \centering
    \includegraphics[width=1\linewidth]{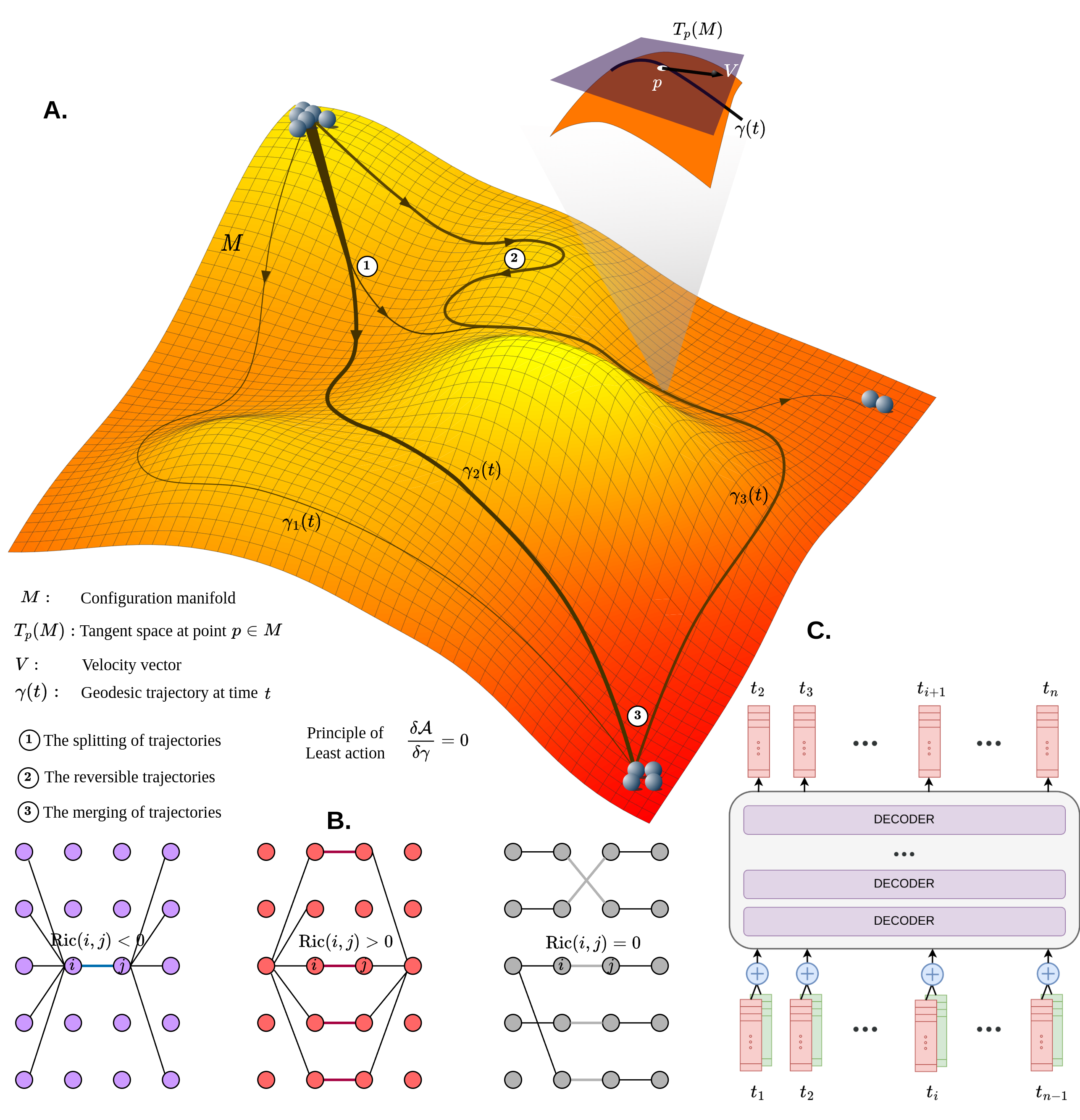}
    \caption{\textbf{A.} The Waddington landscape represented as a high-dimensional manifold ($M$). Different geodesic trajectories are associated with varying probabilities, indicated by the thickness of each trajectory. These trajectories can diverge (at $1$), merge (at $3$), or even locally reverse direction (at $2$). The tangent space $T_p(M)$ is illustrated for an arbitrary point $p$ along the geodesic trajectory $\gamma(t)$. 
    \textbf{B.} Various curvature scenarios in a multipartite graph. In the leftmost panel, negative curvature represents a bottleneck in information flow between nodes $i$ and $j$ which is crucial for passing information between the layers (\textbf{Bridge}). The middle panel illustrates positive curvature, where multiple shortcut paths exist, with $(i,j)$ being just one of them, this shows robustness in information flow (\textbf{Hub}). The rightmost panel depicts zero curvature, indicating a neutral information flow structure. $\textbf{C.}$ The model architecture used in this study is based on an autoregressive transformer model composed of stacked decoder layers. The inputs at each time point include both cell expression data and cell type information, which are summed and used to predict the next cell expression state.}
    \label{fig:landscape}
\end{figure}

\section{Results}
\label{sec:results}
\subsection{Overview of Principle of Least Action}
\label{sec:overview}
The Principle of Least Action is a fundamental law of nature that appears in various contexts within classical and modern physics. It states that, among all possible trajectories a system could follow between two fixed points in configuration space over a given time interval, the actual trajectory is the one that makes the \textbf{action functional}, $\mathcal{A}$ stationary—this trajectory is known as a \textbf{geodesic}. Simply put, $\mathcal{A}(\gamma)$ is a functional that evaluates a given trajectory by computing its associated energy. 

This \textit{variational principle} provides a powerful framework for understanding stochastic dynamical systems including biological processes. In particular, when modeling cellular decision-making, we can extend this concept by considering the epigenetic landscape—often described as Waddington’s landscape—not just as a static energy potential but as a dynamically evolving configuration space. Here, the relevant configuration space, $M$, consists of cells characterized by different gene expression states (Figure \ref{fig:landscape}A). If each cell is represented as $\mathbf{C}=\left(x_1, x_2, \ldots, x_n\right)$ in which $x_i$ is the expression level of gene $i$ collectively, then the system forms a high-dimensional space known as the \textit{gene regulatory network state space} or \textit{epigenetic state space}. This manifold is endowed with a \textbf{metric} or \textbf{connection} that encodes how gene regulation and epigenetic factors shape and constrain a cell’s developmental trajectory. The associated tangent space $TM$ contains the velocity vectors in each position on the manifold. 

The principle of least action, when applied to cellular decision-making, suggests that the developmental trajectory of a cell follows a path that optimally balances regulatory constraints and energy expenditure within the gene regulatory network state space. This naturally leads to the concept of \textit{geodesic flow}, which describes how these optimal trajectories evolve over time. The geodesic flow $G^t(V)$ is the flow on the tangent space $TM$ for the velocity vector $V \in T_p M$ at point $p \in M$. This flow is determined by the derivative of geodesic:

\begin{equation}
    G^t(V):=\dot{\gamma}_V(t)
\end{equation}

The \textbf{geodesic curves} $\gamma(t)$ are the trajectories that minimize the action functional:

\begin{equation}
    \frac{\delta \mathcal{A}}{\delta \gamma}=0
\end{equation}

By applying the \textit{calculus of variations}, we move beyond analyzing discrete points or static forces and instead focus on entire trajectories and how energy is distributed over time. This perspective aligns with the primacy of trajectories discussed in the section \ref{sec:introduction}. It is important to note that we are refer to \textit{geodesic trajectories} rather than a single  optimal trajectory, because there is not a unique optimal solution for the above functional derivative equation.

Since geodesic trajectories minimize the action functional $\mathcal{A}$, their properties are inherently tied to the underlying geometry of the configuration space. To formalize this connection, we express the action functional in terms of the metric tensor $g$, which encodes the geometric structure of the manifold:

\begin{equation}
    \mathcal{A}(\gamma)=\int_0^L \sqrt{g_{\gamma(t)}(\dot{\gamma}(t), \dot{\gamma}(t))} d t.
\end{equation}

the pseudo-Riemannian metric tensor $g_{\gamma(t)}$ determines the geometric structure of the manifold $M$. It determines the distances and angles, and also the curvature of the space, which influences the possible trajectories that cells can take in their developmental process. In section \ref{sec:curvature_analysis} we introduce a surrogate metric, the \textit{Balanced Forman-Ricci curvature} to numerically approximate the the curvature of the configuration manifold (Figure \ref{fig:landscape}B).

\subsection{Computational Reconstruction of Action}
\label{sec:trajectory_prediction}

Understanding the geometric structure of the configuration manifold through the pseudo-Riemannian metric and its curvature provides a robust theoretical foundation for modeling cellular trajectories. However, in many biological contexts, the explicit mathematical form of this configuration space is either unknown or too complex to derive analytically. To address this, we adopt a \textbf{reverse approach}: rather than starting with a known action functional to derive trajectories, we first construct a dataset of observed trajectories and then infer the underlying action functional from this data. This data-driven strategy allows us to approximate the geodesic structure of the system without requiring an explicit analytical model of the configuration space.

More specifically, we employ an \textbf{autoregressive transformer model} \cite{radford2019language} to learn cellular trajectories (Figure \ref{fig:landscape}C). This choice is motivated by the Transformer's ability to capture dependencies across the \textit{entire trajectory}, rather than relying on simplified models based on Markov assumptions. Autoregressive Transformers have proven highly effective in modeling high-dimensional data with intricate temporal dependencies across different data modalities \cite{hong2022cogvideo, ma2024latte, ren2020fastspeech, geneva2022transformers, karniadakis2021physics, nguyen2023scaling}, making them well-suited for capturing the dynamics of time-evolving cellular states.

As demonstrated in Section \ref{sec:formulate_loss_function}, transformer models learn the \textit{paths of least action}, implicitly learning to minimize the action functional approximated from the data. By training the model on experimentally derived datasets that track cellular development (see Section \ref{sec:create_dateset}), we aim to approximate the underlying structure governing cell fate transitions.

To assess the model’s performance and ensure that it effectively reconstructs meaningful trajectories, we evaluate it using three key metrics:

\begin{enumerate}
    \item \textbf{Loss}: Measures the cross-entropy between the model prediction and the trajectories from the evaluation set. We have to minimize this value.
    \item \textbf{Accuracy}: Evaluates the correctness of orders in time steps.  It does so by comparing the predicted sequence of cell states to their actual temporal progression in the dataset.
    \item \textbf{Coverage}: Assesses how comprehensively the model explores the full distribution of cell states, independent of their temporal order. A model with high coverage successfully reconstructs the diversity of cellular phenotypes observed in the dataset
\end{enumerate}

\begin{figure}
    \centering
    \includegraphics[width=1.0\linewidth]{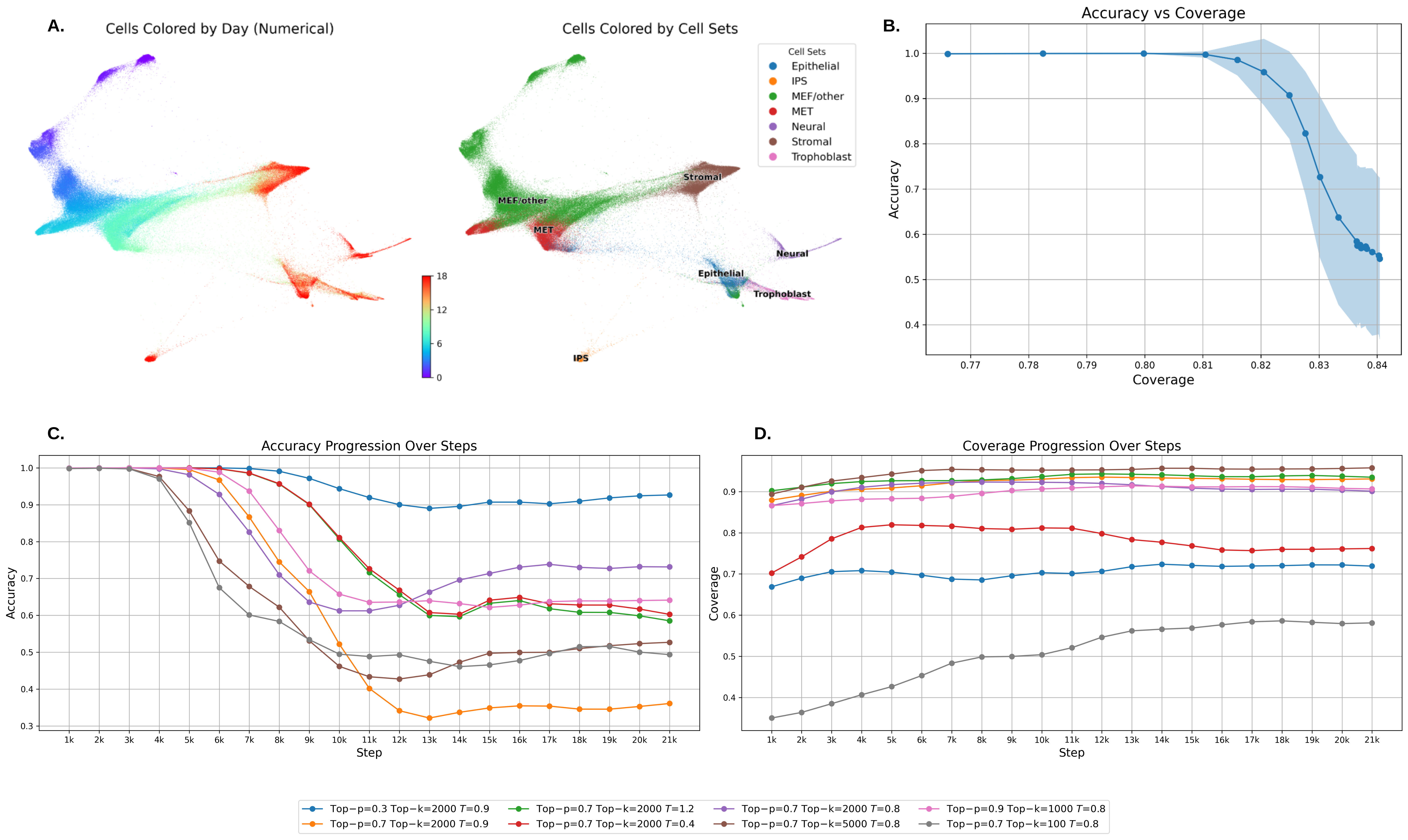}
    \caption{ \textbf{A.} Cell reprogramming dynamics over time and across cell types, visualized using a directed force algorithm. \textbf{B.} Accuracy decreases as coverage increases, but there exists an optimal balance where both accuracy and coverage remain high. \textbf{C., D.} Sample Training Results of the Autoregressive Model. The overall trend indicates that as training progresses and overfitting occurs, accuracy declines while coverage increases. Higher Top$-k$ values and temperature $T$ improve coverage but negatively impact accuracy.}
    \label{fig:accuracy_vs_coverg}
\end{figure}

\begin{figure}
    \centering
    \includegraphics[width=1\linewidth]{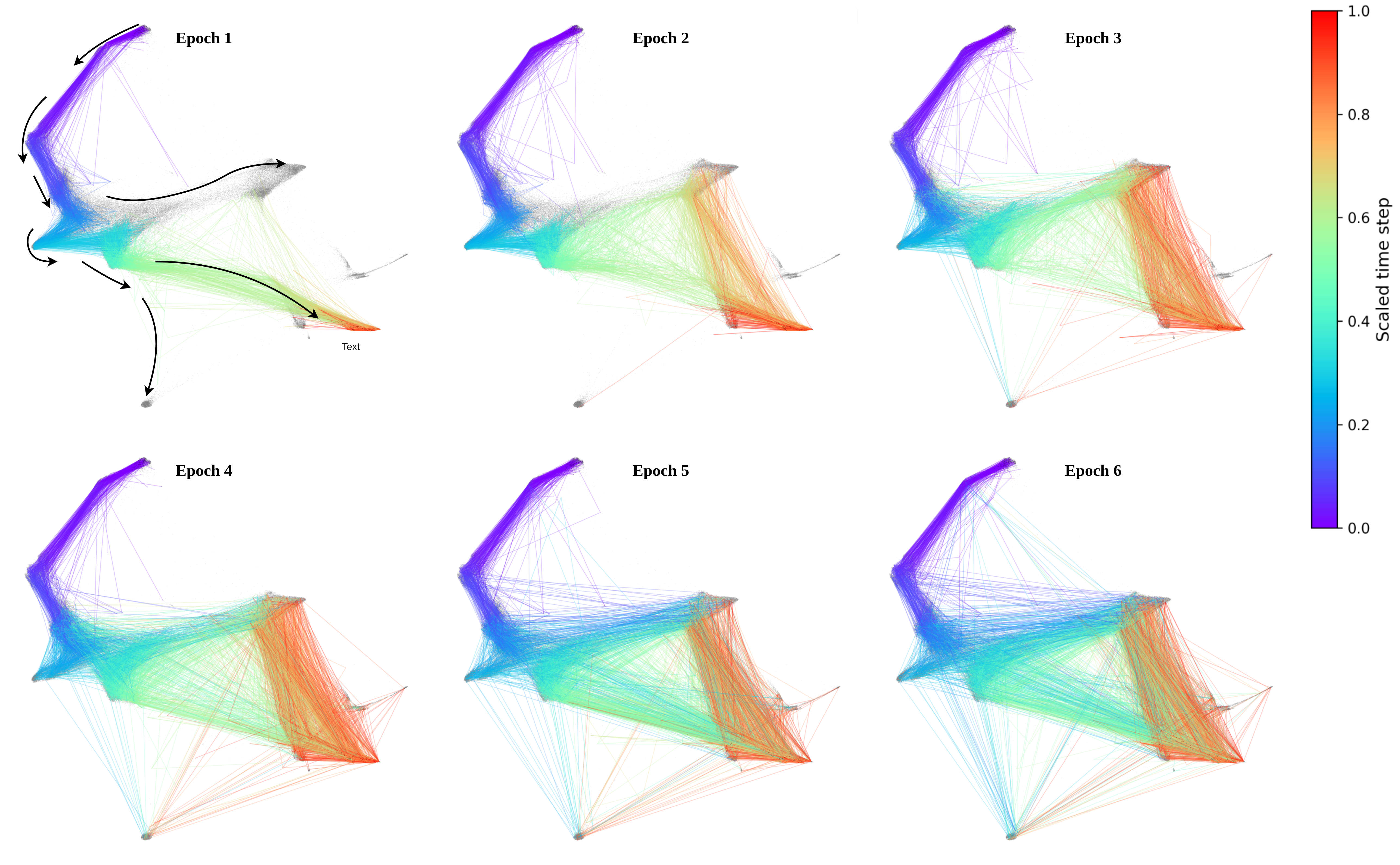}
    \caption{The training of Transformer Autoregressive model for predicting the cell development on evaluation set of Reprogramming of mouse embryonic fibroblasts dataset. As the training progresses the accuracy decreases as the coverage increases}
    \label{fig:training}
\end{figure}

Our experimental results show that the model quickly learns the correct time-step ordering, achieving high accuracy early in training. However, it takes longer for the model to explore the full distribution of possible cell states, leading to a more gradual improvement in coverage (Figure \ref{fig:accuracy_vs_coverg}). As the model overfit to the training set, accuracy tends to decrease while coverage remains relatively stable, highlighting the trade-off between memorizing training data and generalizing to novel trajectories \ref{fig:training}.

Once the model has been trained and evaluated, the next step is to generate predicted cellular trajectories using an appropriate sampling strategy. Since the autoregressive transformer model predicts the next cell state sequentially, the choice of sampling method significantly impacts the quality and diversity of the generated trajectories.

To ensure a balance between diversity and accuracy, we employ \textbf{Nucleus Sampling} \cite{holtzman2019curious} that has three different parameters:

\begin{itemize}
    \item Top$-p$: selects cells from the smallest set whose cumulative probability exceeds a chosen threshold $p$.
    \item Top$-k$: restricts the cell choice to the $k$ most likely cells, discarding all others.
    \item Temperature ($T$): controls the randomness of the sampling of cells, higher temperatures correlate with more random choices
\end{itemize}

We generate trajectories with different values of each parameter: the higher Top-$k$ and Top-$p$ increase the diversity of possible cell choices at each time step, allowing for greater exploration of the state space which comes at the cost of reduced accuracy, as the model may deviate from the most probable trajectories (Figure \ref{fig:accuracy_vs_coverg}C and D). 


To gain a deeper understanding of how trajectories evolve over time, we introduce novel metrics to quantitatively analyze the structure and characteristics of the cellular development manifold. In the next sections we will discuss entropy, curvature, and reversibility as measures that allow us to investigate how cells explore their developmental landscape and identify constraints that govern their transitions.

\subsection{Entropy Measures Exploration-Exploitation Ratio in Development}
\label{sec:entropy_analysis}
Entropy serves as a measure of diversity in the model’s decision-making process at each time step. By examining how entropy evolves over different stages of development, we can gain insights into how the cells balance exploration (diverse trajectory) and exploitation (following the most likely trajectories). During the cellular decision-making process, cells follow specific exploration strategies depending on their context. At each stage, they evaluate both past and current states to compute the probability of transitioning to subsequent states within the phase space of gene expression values. These transitions can be represented as probability distributions over a defined parameter space.

When modeling this process within an autoregressive modeling framework, the probability distribution at each time step can be adjusted using a temperature parameter. In this context, a higher temperature increases randomness, promoting broader exploration of possible states, whereas a lower temperature constrains the model to follow more deterministic pathways, favoring the most probable transitions at each step. This same principle applies to our cellular trajectory predictions, where adjusting the temperature parameter influences the entropy of the system, shaping how the model navigates the developmental landscape.

Since temperature controls the balance between exploration and exploitation in trajectory generation, increasing the temperature generally leads to higher entropy in the system. However, because the number of cells varies at each time step $t$, a direct comparison of entropy values across different time points would be misleading. To address this, we normalize the entropy $H_t$  by dividing it by the logarithm of the number of cells at that time step, $\log(N_t)$ to ensure a fair comparison across stages:

\begin{equation}
    H_{normalized} = \frac{-\sum_{i=1}^{N_t}p_t\log(p_t)}{\log(N_t)} 
\end{equation}

\begin{figure}
    \centering
    \includegraphics[width=1\linewidth]{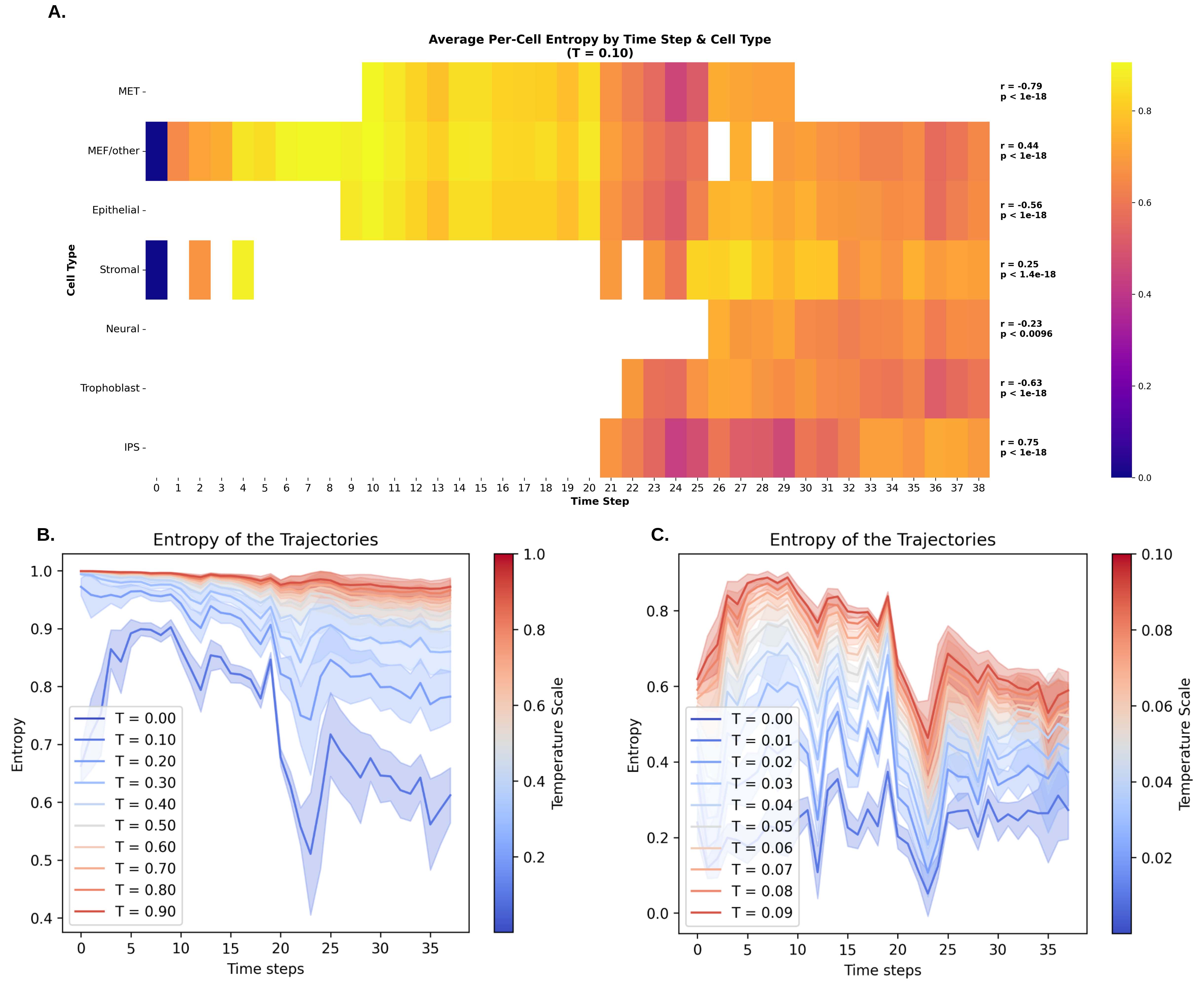}
    \caption{Normalized Entropy Across Different Time Steps
    \textbf{A.} Normalized Entropy of different cell types across time steps at fixed temperature ($T=0.1$): MET is the highest while IPS has the lowest average entropy (Empty boxes indicate missing data due to scarcity).
    \textbf{B.} Entropy fluctuations on the temperature scale $[0, 1]$. Higher temperatures smooth out differences in next-cell selection, while lower temperatures make these differences more pronounced. \textbf{C.} A zoomed-in view of normalized entropy in the $[0, 0.1]$ range, revealing finer details of fluctuations. Notably, two valleys appear at time steps 12 and 24.}
    \label{fig:entropy_analysis}
\end{figure}
The normalized entropy $H_{normalized}$ provides a measure of how extensively the developmental process explores the expression phase space at different stages. Figure \ref{fig:entropy_analysis} presents the entropy of trajectories averaged over 100 samples, computed separately for different temperature values.

A key trend observed in the entropy analysis is a gradual decrease in entropy as cells transition from less differentiated states to their final stages, consistent with findings reported in Baptista et al. \cite{baptista2024charting}. However, this decline is not uniform throughout the trajectory of reprogramming dataset. A sharp drop in entropy occurs at time step 24 (corresponding to day 11, 11.5 days post-Dox withdrawal), after which entropy rises again in subsequent time steps. This increase could suggest that cells begin to explore a significantly larger expression space following a period of constraint, representing a bottleneck in the developmental process. Interestingly, this entropy pattern aligns with the curvature dynamics observed in Figure \ref{fig:curvature_analysis} that we discuss further in the next section. 

A more detailed view emerges when the temperature is lowered, allowing us to capture the deterministic aspects of each time step through cell-cell similarity dynamics. Another notable pattern in Figure \ref{fig:entropy_analysis}A is that entropy levels vary across cell types in reprogramming dataset. MET (mesenchymal-to-epithelial transition) and MEF/other cells (mouse embryonic fibroblasts) exhibit the highest entropy, particularly during the early stages of development, indicating broad exploratory behavior. In contrast, IPS (Induced Pluripotent Stem) and Trophoblast cells maintain the lowest average entropy which indicates a more deterministic and constrained differentiation trajectory.

While entropy provides a statistical view of how broadly cells explore the expression phase space during development, it does not fully capture the geometric structure that shapes these trajectories. To complement the entropy analysis, we turn to the study of curvature, which offers a deeper understanding of the underlying constraints and dynamical tendencies of the configuration manifold.

\subsection{Curvature Quantifies Information Flow}
\label{sec:curvature_analysis}

By viewing cell fate as a process, an ensemble of cells moving along geodesics in high-dimensional configuration space, we can apply tools from differential and information geometry \cite{amari2016information} to analyze how developmental information propagates over time. These geometric approaches provide insight into how information flows for long-distance message passing \cite{topping2021understanding}, especially for systems far from equilibrium, where traditional equilibrium thermodynamic models fall short.

In particular, Riemannian geometry has been used in non-equilibrium thermodynamics to study system behavior through concepts such as Ruppeiner geometry \cite{ruppeiner1979thermodynamics, ruppeiner1995riemannian}, which has successfully explained phenomena in complex, non-Boltzmannian dynamics \cite{crooks2007measuring, aaman2008thermodynamic}. Although the metric tensor $g$ encodes the full local structure of the manifold, in practice, we often focus on curvature, a scalar quantity that captures the global geometric properties and the directional behavior of trajectories. Recent studies have extended this framework to biological systems, employing discrete analogs of curvature to study genomic and regulatory network data \cite{pouryahya2017comparing, murgas2022quantifying}, and to explore its relationship with entropy and information flow \cite{sandhu2015graph, murgas2022hypergraph}. In our context, curvature analysis allows us to interpret how developmental constraints and regulatory architecture shape the space through which cells transition, providing a complementary lens to both trajectory prediction and entropy dynamics.

We use the learned cell embeddings to construct a \textbf{multipartite graph} $G=(V, E)$ to approximate the curvature. Like manifolds, graphs exhibit curvature that makes them appropriate for studying spaces with hyperbolic geometry \cite{liu2019hyperbolic, boguna2021network}. In this graph, the set $V$ represents the cells, which we organize into $k=N$ layers, where $N$ corresponds to the number of time steps. An edge $(i, i+1) \in E$ exists if and only if the two cells are consecutive in a trajectory $\gamma$, meaning $C_i, C_{i+1} \in \gamma$ for some $\gamma$. The degree of the cell $i$, denoted by $d_i$,  represents the total number of trajectories entering or exiting the cell.

Following \cite{topping2021understanding}, we use the \textbf{Balanced Forman curvature}. This choice is primarily because of the limitations of Forman \cite{forman2003discrete} and Ollivier \cite{ollivier2007ricci, ollivier2009ricci} curvatures, which tend to be either skewed toward negative values and pose challenges in controlling local quantities \cite{munch2023non}.

First, we need the following definitions:
\begin{itemize}
    \item $T_{i,j}$: number of triangles based on $i \sim j$
    \item $S_i$: number of 4-cycles without inside diagonal at $i$
    \item $\gamma_{\max }(i, j)$: Maximal number of 4-cycles based on $i \sim j$ traversing a common node 
\end{itemize}
The Balanced Forman curvature for an edge $i \sim j$ is defined as follows:
\begin{equation}
    \operatorname{Ric}(i, j)=\frac{2}{d_i}+\frac{2}{d_j}-2+2 \frac{T_{i j}}{\max \left\{d_i, d_j\right\}}+\frac{T_{i j}}{\min \left\{d_i, d_j\right\}}+\frac{S_i+S_j}{\gamma_{\max } \max \left\{d_i, d_j\right\}}
\end{equation}

The $\operatorname{Ric}(i, j)=0$ when $\min \left\{d_i, d_j\right\}=1$. Unlike the Ollivier or Forman curvature that are biased towards negative curvature, this definition gives a more balanced values to curvature across the graph.

In this context, two types of connections can be distinguished:

\begin{itemize}
    \item \textbf{Bridges (Negative curvature)}: These edges function as bridges or bottlenecks for information. Removing them increases the average shortest path length in the graph. Their role is to map information to a lower-dimensional space, a phenomenon known as \textit{oversquashing} \cite{topping2021understanding}.
    \item \textbf{Hubs (Positive curvature)}: These edges serve as backup trajectories, and their removal does not alter the shortest paths. Their primary function is to enhance the robustness of the network.  
\end{itemize}

The above definitions are also reflected in \cite{elkin2024dynamic}, where in gene regulatory or co-expression networks, positive Ricci curvature often signifies a tightly knit cluster of genes (many redundant connections and feedback loops), whereas negative curvature marks sparse bridges between modules. Recently, the curvature has been used for calculating the \textit{network robustness} in biological system that captures global gene signaling changes in functional cooperation \cite{simhal2024orco}.

In order to calculate the curvature for each node (cell), the curvature of each node is calculated based on the average of in-coming and out-going edges:

\begin{equation}
    \kappa(i)=\frac{1}{d_i} \sum_{i:(i, j) \in E} \operatorname{Ric}(i, j)
\end{equation}

\begin{figure}
    \centering
    \includegraphics[width=1\linewidth]{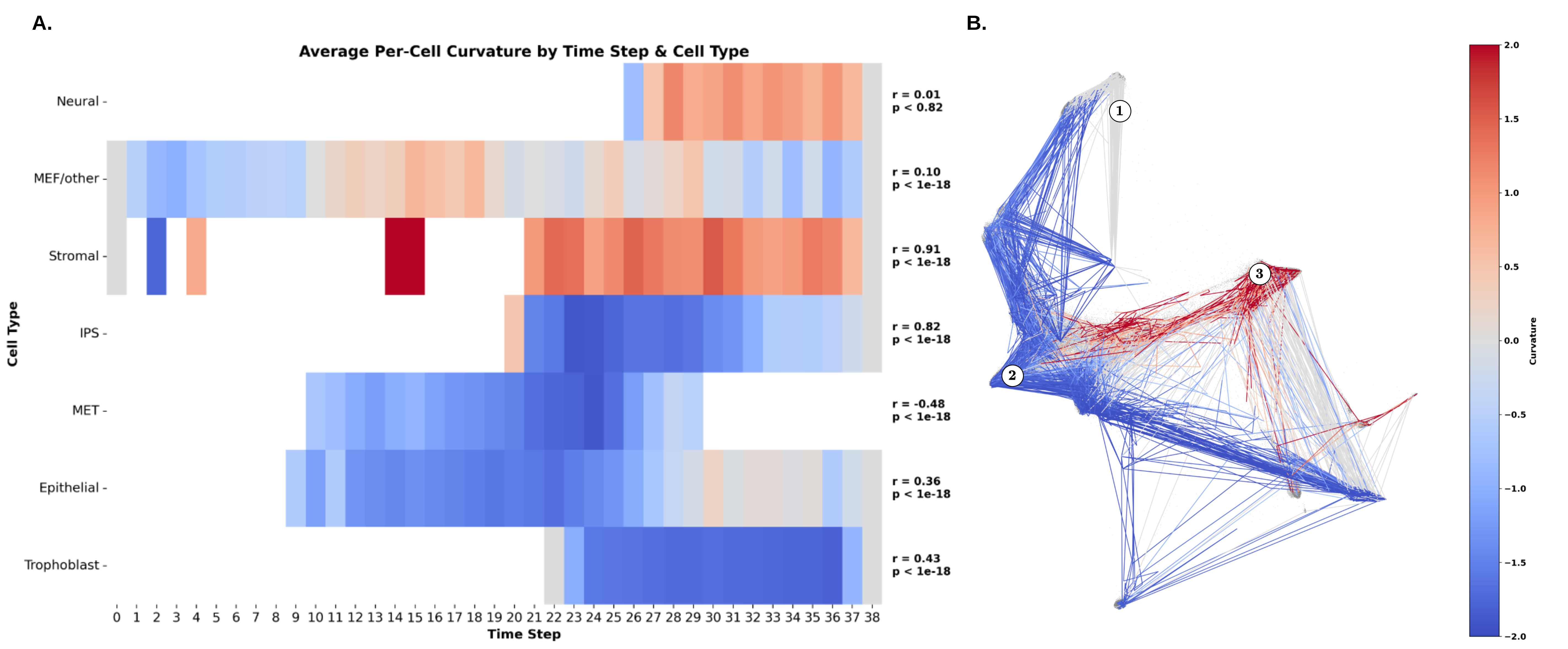}
    \caption{Curvature Analysis of Cell Development in the Evaluation Set of Mouse Embryonic Reprogramming. \textbf{A.} The average curvature across time steps and cell types reveals distinct differences between both cell types and developmental stages. \textbf{B.} Cell development trajectories are color-coded based on Balanced Forman curvature. Initially, curvature remains mostly flat ($1$). In most trajectories, particularly during intermediate time steps ($2$), curvature becomes negative, indicating the \textbf{oversquashing} phenomenon, characteristic of \textbf{Bridges}. In contrast, regions with positive curvature ($3$) predominantly appear in \textit{Stromal} and \textit{Neural} cell types, suggesting excessive connectivity, characteristic of \textbf{Hubs}.}
    \label{fig:curvature_analysis}
\end{figure}

As we can see from the Figure \ref{fig:curvature_analysis}, the curvature follows very different dynamics for different cell types. Specifically, the curvature for Neural, mouse embryonic fibroblasts (MEF)/Other, and Stromal cell types is more positive, an indication that they pass through network hubs with feedback loops and redundant signaling pathways, which support promiscuous signaling and robust information flow \cite{banerji2013cellular}. This means the information flow of gene expression in different time steps are more robust, and even removing the signalings will not compromise the differentiation process as much as the other cell types. \textit{Neural cells} and \textit{stromal} fibroblasts, which respond to a wide range of developmental and environmental cues, often overexpress network hub genes to distribute signaling efficiently across the genome \cite{teschendorff2021ultra, shi2020quantifying, teschendorff2017single}. Such overexpression of hub genes effectively ``draws in'' signaling flux and spreads it over a highly interconnected network, increasing the efficiency and robustness of signal propagation. For instance, Sandhu and colleagues observed globally increased network curvature in stem-like cells (which include pluripotent cells and, by extension, plastic stromal cells) \cite{murgas2022hypergraph}. In practical terms, a \textit{neural} or \textit{mesenchymal} cell’s fate decisions are buffered by this robust network: multiple signals and feedback loops ensure that development proceeds smoothly even in noisy conditions.

In contrast, for \textit{IPS} (induced pluripotent stemcells), \textit{MET} (mesenchymal-to-epithelial transition), epithelial and trophoblast cell types are specialized in passing the necessary information by a lower dimensional representation over \textbf{Bridges}. This is crucial for special roles these cell types have in their final state. These systems have gene regulatory connections that are more tree-like or linear, indicating specialized pathways with less redundancy.  during \textit{iPSC reprogramming} of fibroblasts, cells must pass through a transitional \textit{MET} state that is particularly fragile – blocking the \textit{MET} gene program (e.g. E-cadherin induction) aborts reprogramming \cite{li2010mesenchymal}. The \textit{MET} stage represents a narrow bridge between \textit{mesenchymal} and \textit{pluripotent} networks; its gene expression dimensionality is low and easily perturbed, consistent with negative curvature edges connecting the \textit{fibroblast} module to the \textit{pluripotency} module.

Curvature analysis reflects this phenomenon through an abundance of low or negative curvature edges, corresponding to bridge-like connections with limited information exchange \cite{elkin2024dynamic}. In \cite{banerji2013cellular}, it's been demonstrated that signal promiscuity diminishes in favor of a few dominant pathways, which leads to gene expression patterns that become increasingly canalized and exhibit lower curvature. Similarly, findings in \cite{baptista2024charting} confirm that total Ricci curvature decreases in accordance with our observations of curvature flattening as cells approach the terminal stages of differentiation.

\subsection{Local Reversibility Is A Metric of Developmental Asymmetry}
\label{sec:local_irreversibility}
One of the most fundamental aspects of biological physics is the concept of irreversibility \cite{england2015dissipative, collin2005verification, bustamante2021optical}. This principle, rooted in the second law of thermodynamics, was also recognized early in biology through \textbf{Dollo's law of irreversibility} \cite{louis1893lois}, which states ``an organism never fully return to its initial condition in which it has previously lived''. Despite the centrality of this concept in biological development, many current approaches \cite{schiebinger2019optimal, shi2019quantifying, zhou2021dissecting} disregard irreversibility by treating the Waddington landscape as a conservative potential field function, $V(x)$, where traversing a closed path results in no net change to the system’s state which is the hallmark of reversible dynamics. Similarly, some models describe statistical dynamics in equilibrium with steady-state conditions, incorporating deterministic drift and random forces via \textit{Langevin dynamics}. However, such models assume the system is near equilibrium with \textbf{detailed balance}, a condition that fails in open systems with energy flux and in active matter or non-Boltzmann conditions, where reversibility is broken \cite{o2022time}.

The concept of irreversible differentiation was challenged—and eventually overturned—by pivotal experiments in the 20th and 21st centuries. These studies demonstrated that, under appropriate conditions, a cell's developmental trajectory can be reversed or reset. In 1962, John Gurdon’s work on amphibian nuclear transfer provided the first clear evidence that differentiated cells retain the genomic information necessary to drive embryonic development. His experiments showed that the nucleus of a specialized intestinal cell could be ``reprogrammed'' by an enucleated egg to form a tadpole, establishing that genomic potential is preserved during differentiation \cite{gurdon2006nuclear}. Since then, cellular reprogramming, transdifferentiation, and even dedifferentiation have been observed, highlighting that cell fate is experimentally malleable \cite{loi2013sheep, kalra2021cell}.

These findings suggest that biological development is neither fully reversible nor completely irreversible but rather a complex interplay of both. In this section, we introduce a generalized framework that quantifies ``local (ir)reversibility'' by employing the concept of \textbf{local entropy production}. This measure not only indicates the intensity of reversibility but also pinpoints its location within the process of cellular development.

Suppose a trajectory as follows:
\begin{equation}
    \gamma = (\mathbf{x}_0, \cdots, \mathbf{x}_{T})
\end{equation}
we can also define the reverse of this trajectory:
\begin{equation}
    \tilde{\gamma} = (\mathbf{x}_{T}, \cdots, \mathbf{x}_{0})
\end{equation}

In equilibrium, the system satisfies \textbf{detailed balance}, meaning the probability of any forward trajectory is exactly equal to that of its reverse. In contrast, \textbf{nonequilibrium} systems violate detailed balance,  leading to a difference between the forward and reverse trajectory probabilities.
\textbf{Entropy production} quantifies how far a system deviates from detailed balance. In other words, entropy production is a measure of \textit{time-reversal symmetry breaking}. As dictated by the second law of thermodynamics, entropy production is always non-negative \cite{mandal2017entropy}. For a given trajectory $\gamma$, the entropy production is defined as:
\begin{equation}
    \sigma(\gamma)=k_B \log \frac{P_\theta(\gamma)}{P_\theta(\tilde{\gamma})}
\end{equation}

\begin{figure}[H]
    \centering
    \includegraphics[width=1\linewidth]{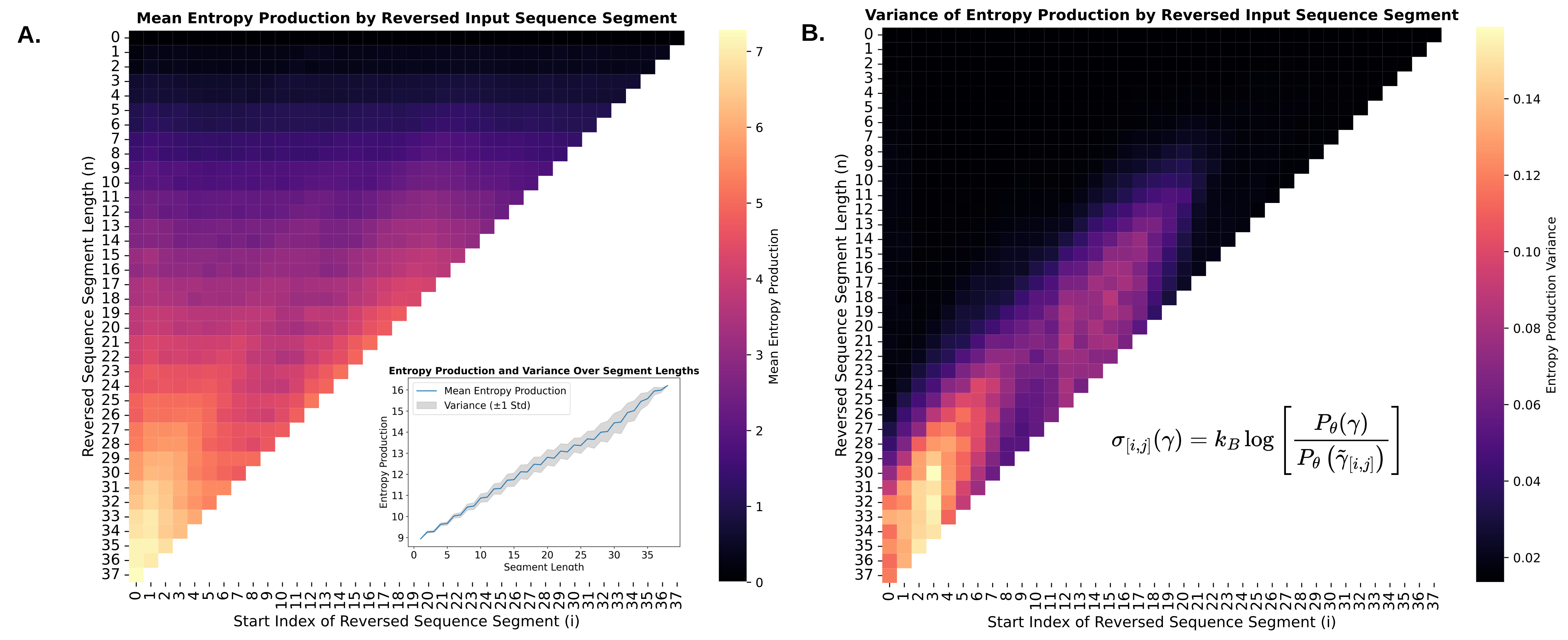}
    \caption{Mean and Variance of \textbf{Entropy Production} Across Different Reversed Trajectories. \textbf{A.}  Entropy production increases with longer sequence segments. \textbf{B.} The variance in entropy production remains generally low but increases for longer sequence segments.}
    \label{fig:entropy_production}
\end{figure}

where $k_B$ is the Boltzmann constant. When the system satisfies detailed balance or is driven by a conservative field, the entropy production is zero. The \textbf{Kullback-Leibler (KL) divergence} between the \textit{forward trajectory probability distribution} $P_\theta(\gamma)$ and the \textit{backward trajectory probability distribution}  $P_\theta(\tilde{\gamma})$ is given by:
\begin{equation}
    D_{\mathrm{KL}}(P_\theta(\gamma) \| P_\theta(\tilde{\gamma}))=\sum_\gamma {P_\theta}(\gamma) \log \frac{{P_\theta}(\gamma)}{{P_\theta}(\tilde{\gamma})}
\end{equation}

Now we can calculate the \textit{expected entropy production} over an ensemble of trajectories as follows:
\begin{equation}
    \left\langle\sigma(\gamma)\right\rangle_{\gamma \in \Gamma'}=k_B \sum_{\gamma \in \Gamma'} {P_\theta}(\gamma) \log \frac{{P_\theta}(\gamma)}{{P_\theta}(\tilde{\gamma})}
\end{equation}

Unlike most previous approaches \cite{shi2020quantifying, huang2017processes} that assume detailed balance, we demonstrate that development is predominantly an irreversible process—forward trajectories are far more likely than their reversals. Nonetheless, it is important to recognize that developmental trajectories can occasionally involve a ``dedifferentiation'' process, whereby a cell reverts from a partially or terminally differentiated state to a less differentiated one within the same lineage \cite{yao2020dedifferentiation}. This phenomenon provides evidence that these biological processes can exhibit \textit{local reversibility} under certain conditions. Modeling such processes as either fully reversible or completely irreversible fails to capture these nuances. To address this, we generalize the concept of entropy production to account for locally reversible trajectories, allowing us to analyze where reversibility is most likely to occur during development.

To achieve this, we define a \textbf{locally time-reversed trajectory} for the interval $[i, j]$ ($i<j$) as follows:
\begin{equation}
    \tilde{\gamma}_{[i, j]}=(\mathbf{x}_0, \cdots, \mathbf{x}_j, \cdots \mathbf{x}_i, \cdots , \mathbf{x}_T)
\end{equation}

In this definition, the time intervals where $t<i$ or $t>j$ remains identical to those of the original trajectory, while for $i \leq t \leq j$ is reversed. The \textbf{local entropy production} is defined as follows:
\begin{equation}
    \sigma_{[i, j]}(\gamma)=k_B \log \left[\frac{P_\theta(\gamma)}{P_\theta\left(\tilde{\gamma}_{[i, j]}\right)}\right]
\end{equation}

Given the learned model $P_\theta$, the local entropy production has been calculated across different time steps as shown in Figure \ref{fig:entropy_production}. In \textbf{panel A}, the x-axis represents the starting point of the reversed sequence segment ($i$), while the y-axis indicates the segment length ($j-i$). As we move downward in the heatmap, segment lengths increase, leading to higher entropy production, as expected. However, the figure also reveals additional interesting patterns. Notably, entropy production is negligible for segments shorter than $5$, suggesting that the cell development process is locally reversible at small segment sizes. This observation aligns with the behavior of many cells that exhibit local oscillations within the gene expression manifold, without committing to an irreversible change in fate, consistent with previously reported gene expression dynamics \cite{leng2015oscope}. Computational studies of gene regulatory circuits also recapitulated this behavior \cite{ramirez2024dissecting}, consistent with the local reversibility feature. Conversely, larger segments exhibit more pronounced irreversibility at later time points, and the maximum entropy production corresponds to reversing the total process.

It's important to note that this observed pattern is not uniform throughout the process. For example, a darker band appears between time steps 6 and 9, corresponding to segment sizes roughly between 19 and 30, suggesting that cell trajectories are more likely to retrace their trajectories at later stages than at earlier ones. As shown in \cite{wang2010potential}, strong asymmetric potential barriers inherently favor the forward direction. Additionally, an epithelial-to-mesenchymal transition (\textit{EMT}) study found that once cells have transitioned, the reversal (mesenchymal-to-epithelial) is more likely to occur at later time points but not fully terminal states. At the same time the variance of entropy production is generally low (Figure \ref{fig:entropy_production} panel B) but increases with larger segment sizes, indicating greater variability in the reversibility of cell development over longer trajectories and at later time points.


\section{Discussion}
\label{sec:discussion}
Traditionally, cell and developmental biology have long been viewed as distinct disciplines differentiated by their specific research focus. However, with the advent of molecular biology techniques, it has become increasingly evident that both fields fundamentally seek to understand similar biological processes \cite{dawes2001cell}. In general, insights into how cells develop and specialize inform key phenomena such as differentiation, dedifferentiation, embryogenesis, and tissue homeostasis.

Despite significant progress in measurement techniques, the overarching principles that guide cellular development remain qualitative and only partially understood. Here, we demonstrate that cell fate dynamics conform to the principle of least action and introduce a computational framework capable of quantifying action from time-series data points. Our approach has several advantages:

\begin{enumerate}
    \item As opposed to interpreting the Waddington landscape as a conservative potential function, where movements along a closed path do not result in net change \cite{schiebinger2019optimal, shi2019quantifying, zhou2021dissecting}, we incorporate irreversibility explicitly in the trajectories of cell development, reflecting their inherent time-asymmetry.

    \item Many current modeling approaches assume the system is in equilibrium or a steady state, and describe its behavior using Langevin dynamics, which combine deterministic drift with random forces modeled as Gaussian noise \cite{o2022time,cao2025stochastic}. In contrast, our approach is tailored for systems operating far from equilibrium, a more appropriate framework for biological processes, which are inherently directional and dissipative

    \item Our method does not rely on the Markov assumption, which treats cell state transitions as memoryless. Instead, we allow for history-dependent transitions, capturing potential affinities and path dependencies between states across time.

    \item We avoid averaging over heterogeneous cell trajectories, which can blur distinct patterns in cell behavior. By focusing on the least action path for each trajectory, our method preserves the fine-grained structure of cell state evolution, that can't be captured by a general Jacobian field.
    
    \item Our framework is grounded in well-established physical principles, providing a more rigorous and interpretable foundation for understanding cell decision-making. By anchoring developmental dynamics in the variational principle of least action, we open up a path for integrating tools from statistical physics and dynamical systems theory with modern single-cell measurement techniques.
\end{enumerate}

Furthermore, we propose entropy, curvature, and irreversibility as key quantitative metrics for understanding how cells commit to specific developmental pathways. High entropy levels reflect extensive exploration of the gene expression landscape—a signature of cellular plasticity. As cells commit to defined lineages, entropy decreases, signaling a loss of plasticity and a shift toward more deterministic behavior. The degree of commitment, plasticity, can be further quantified by irreversibility: higher irreversibility, as measured by our proposed metric, indicates greater plasticity and the capacity for dynamic transitions. Concurrently, our curvature analysis reveals that regions of positive curvature correspond to robust, redundant signaling hubs—'elastic zones' that can buffer perturbations. Conversely, regions with negative curvature delineate critical bottlenecks or bridges, indicating areas where the system is less resilient, more fragile, and susceptible to disruption.

Despite these advances, several limitations merit consideration. First, our approach is dependent on the availability of high-resolution, time-series single-cell data; noise and sampling biases in such datasets may compromise trajectory reconstruction. Second, while transformer-based architectures effectively capture long-range dependencies, their high computational cost could limit scalability to even larger or more complex datasets. Finally, although entropy and curvature provide powerful quantitative insights, the biological interpretation of these abstract metrics requires further experimental validation to fully substantiate their relevance in vivo.

By bridging physical principles with biological complexity, our computational framework provides a novel perspective on cellular development and pathological deviations. This approach opens new avenues for future research by enabling data-driven analyses across diverse biological datasets and emphasizes the potential of machine learning to capture complex temporal dynamics, ultimately advancing the development of more predictive and interpretable models of biological systems.



\section{Methods}
\label{sec:methods}

Current methods for modeling scRNA dynamics rely on arbitrary, restrictive assumptions that hinder the generalization of these dynamics and the handling of out-of-distribution cases. Rather than treating the Waddington landscape \cite{schiebinger2019optimal, shi2019quantifying, zhou2021dissecting} as a conservative potential function, we need a model that accurately captures irreversibility. This requires accounting for the \textbf{non-equilibrium} dynamics that are fundamental to biological processes, as emphasized by pioneers such as Prigogine \cite{prigogine1982being, prigogine2018order} and Lotka \cite{lotka1956elements}. To address these challenges, our model should also move beyond the traditional Markov assumption to incorporate memory effects in developmental processes.

Some approaches \cite{qiu2022mapping} rely on ``averages'' over trajectories that originate from and lead to various points in cell history, thereby overlooking the detailed individual cell movements that cannot be captured by a general ``Jacobian field.''

The issues mentioned above arise from arbitrary modeling with weak physical foundations, which fail to provide a comprehensive framework for cell differentiation dynamics. We propose a model that integrates concepts from physics and information theory, leveraging data-driven techniques from modern machine learning to bridge this gap and offer deeper insights into the underlying dynamics.

In this section we lay out our approach which is based on the \textit{principle of least action} that is generalized to a probabilistic setting. We then show how this can be calculated using the standard language model objective.

\subsection{Least Action Principle}
\label{sec:least_action_principle}
In a probabilistic setting, it can be shown that the principle of least action is equivalent to the principle of maximum entropy (MaxEnt) \cite{brissaud2007maxent}. MaxEnt has been proposed as an underlying principle to numerous fields such as ecology \cite{harte2011maximum}, neuroscience \cite{xu2017dynamical}, machine learning \cite{zheng2017understanding} and economy \cite{scharfenaker2020maximum}. Within this framework, there is not a unique trajectory from the initial to the final state but an ensemble of them. Each cell follows a trajectory with a corresponding probability. This trajectory is determined by not only the initial conditions but also the paracrine signaling between cells as well as the randomness of the environment that affects fates.

Assume we have $n$ genes for each cell at time $t$, represented by $n-$dimensional vector $\mathbf{x}_{t}$. Also, the gene velocity is given by $\dot{\mathbf{x}}_{t}$ 
Although time is continuous, our observations consist of discrete samples. Each trajectory $\gamma^i \in \Gamma$ is a sequence of cell states from start (progenitor cells) to target (differentiated or terminal cells), and can be represented as follows:

\begin{equation}
    \gamma^i = (\mathbf{x}^i_0, \cdots, \mathbf{x}^i_{T})
\end{equation}

In any system, to model the \textit{Action}, we need to calculate the Lagrangian of the system. For each trajectory $i$ from time $0$ to $L$ the Action for trajectory $\gamma^i$ is defined as:

\begin{equation}
    \mathcal{A}_\theta(\gamma^i)=\int_{0}^{L} \mathcal{L}_\theta^i(t, \mathbf{x}_t^i, \dot{\mathbf{x}}_t^i) \mathrm{d} t
\end{equation}

where $\mathcal{L}_\theta^i(t, \mathbf{x}^i, \dot{\mathbf{x}}^i)$ is the Lagrangian of the trajectory $\gamma^i$ which is function of expression $\textbf{x}_t^i$ and velocity $\dot{\textbf{x}}_t^i$. We used the parameter $\theta$ to show that the Action functional and the Lagrangian are both unknown but belong to a family of functions parametrized by $\theta$.

For conservative systems, the focus is typically on the ``least action path,'' representing the trajectory with the minimum action. However, in our context, the system explores an ensemble of trajectories rather than a single one. While the Lagrangian can be analytically derived from kinetic and potential energy in simple systems with few degrees of freedom and linear dynamics, this approach becomes infeasible for more complex systems. Instead, we learn the Lagrangian directly from data using a Transformer neural network \cite{cranmer2020lagrangian, vaswani2017attention}. To achieve this, we adopt an alternative formulation of the Lagrangian, leveraging information theory in conjunction with neural networks.

If we consider a family of \textit{statistical models} $p_\theta$ parameterized by $\theta$ to model the distributions of trajectories, such that $p_\theta(\gamma^i)$ represents the likelihood of the trajectory $\gamma^i$, then the expectation of the action over a subset of possible trajectories $\Gamma' \subseteq \Gamma$ can be calculated as:

\begin{equation}
    \mathcal{A}_\theta(\Gamma')=\sum_{\gamma^i \in \Gamma'} p_\theta(\gamma^i) \mathcal{A}_\theta(\gamma^i)
\end{equation}
this is in parallel with the Shannon entropy for trajectories in $\Gamma'$:

\begin{equation}
    H_\theta(\Gamma')=-\sum_{\gamma^i \in \Gamma'} p_\theta(\gamma^i) \log p_\theta(\gamma^i)
\end{equation}

in this context, $H_\theta(\Gamma')$ is the \textit{trajectories entropy} which quantifies the uncertainty associated with predicting the trajectory of a cell in an ensemble of cells that transitioning from time $0$ (progenitor cells) to $L$ (differentiated or terminal cells). This entropy captures the stochastic nature of the developmental process, reflecting variability in possible cellular pathways.

Following \cite{wang2005maximum, wang2006maximum}, we can define the action of a trajectory as follows:

\begin{equation}
    \mathcal{A}_\theta(\gamma^i) = -\log p_\theta(\gamma^i)
\end{equation}
This interpretation suggests that action is equivalent to \textit{Surprise}. This equation provides a fascinating bridge between physics (specifically the principle of least action) and information theory. From a probabilistic perspective, the principle of least action corresponds to selecting the paths that minimizes the negative log-likelihood, $-\log p_\theta(\gamma^i)$. Thus, minimizing the action corresponds to minimizing uncertainty $H_\theta(\Gamma')$ over the possible cellular trajectories $\Gamma'$.

\subsection{Training using a Language Model}
\label{sec:training_lm}
\subsubsection{Creating the Dataset}
\label{sec:create_dateset}
We take a data-driven approach by first extracting possible trajectories from the source to destination cells using scRNA cell expression and velocity data. This allows us to create a dataset of trajectory time sequences. Consider the trajectory dataset comprising $N$ trajectories:

\begin{equation}
    \Gamma' = \{\mathbf{C}_1^i, \cdots, \mathbf{C}_{T_i}^i \} = \{(\{\mathbf{x}_1^i, \dot{\mathbf{x}}_1^i\}, \{\mathbf{x}_2^i, \dot{\mathbf{x}}_2^i\}, \dots, \{\mathbf{x}_{L_i}^i, \dot{\mathbf{x}}_{L_i}^i\}) \:\: \text{for all } i=1:N\}
\end{equation}

in which $x_j^i$ and $\dot{x}_j^i$ are the expression and velocity of the cell at time $j$ in trajectory $i$ respectively and $L_i$ is the length of trajectory $i$, usually all the trajctories have the same length hence $L_i=L$. We can represent all the data we have about each cell at time $j$ as $\mathbf{C}_j^i=\{\mathbf{x}_j^i, \dot{\mathbf{x}}_j^i\}$, this generalized coordinate is also known as \textbf{configuration space}, this can include other cell data such as phenotypes too.

Because scRNA-seq destroys cells in the process of recording them through time, it is impossible to follow the expressions through time. The current approaches to find the cell transitions are mostly rely on finding the correlations between the cells at different times $j$ and $k$, $C_j$ and $C_k$ in a Markov setting as follows \cite{lange2022cellrank, bergen2020generalizing, maizels2023deep}:

\begin{equation}
    p_{\mathrm{j}, \mathrm{k}}=\frac{\exp \left(\operatorname{corr}\left(\mathbf{C}_{\mathrm{j}}, \mathbf{C}_{\mathrm{k}}\right)\right)}{\sum_{l \in \mathcal{N}_{\mathrm{j}}} \exp \left(\operatorname{corr}\left(\mathbf{C}_{\mathrm{j}}, \mathbf{C}_{\mathrm{l}}\right)\right)}
\end{equation}
in which $\mathcal{N}_{\mathrm{j}}$ are the neighbors of cell at time $j$ that was calculated using a KNN (k-nearest neighbor) approach.

To account for the influence of past states in predicting the next state $k$, we can construct a weighted combination of their correlations with the candidate state. A widely used approach is to apply an exponential decay factor $\lambda \in(0,1)$, which ensures that more recent states exert a stronger influence while older states contribute progressively less.

Under this formulation, the influence of a state visited $m$ steps in the past is weighted by $\lambda^m$ leading to a memory-aware transition model where past states contribute in a decaying fashion:

\begin{equation}
    p_{j, k}^{(t)}=\frac{\exp \left(\sum_{m=0}^{L-1} \lambda^{m} \operatorname{corr}\left(\mathbf{C}_{j_{t-1-m}}, \mathbf{C}_k\right)\right)}{\sum_{l \in \mathcal{N}_j} \exp \left(\sum_{m=0}^{L-1} \lambda^{m} \operatorname{corr}\left(\mathbf{C}_{j_{t-1-m}}, \mathbf{C}_l\right)\right)} .
\end{equation}

The non-Markovian transition probability accounts for a weighted history of previously visited states, allowing past correlations to influence the current transition choice.  this approach captures long-range dependencies in the system, making it particularly well-suited for modeling biological processes such as cell differentiation, where past states can significantly impact future transitions.

\subsubsection{Formulating the Loss Function}
\label{sec:formulate_loss_function}
The probability of each trajectory can be represented using the chain rule in its most general form as follows:

\begin{equation}    p_\theta(\gamma^i)=p_\theta\left(\mathbf{C}^i\right)=p_\theta\left(\mathbf{C}_1^i\right) p_\theta\left(\mathbf{C}_2^i \mid \mathbf{C}_1^i\right) \ldots p_\theta\left(\mathbf{C}_{L_i}^i | \mathbf{C}_1^i \mathbf{C}_2^i \ldots, \mathbf{C}_{L_i-1}^i\right)= \prod_i^{L_i} p_\theta\left(\mathbf{C}_t^i \mid \mathbf{C}_{<t}^i\right)
\end{equation}

The action for each trajectory is calculated as:
\begin{equation}
    \mathcal{A}_\theta(\gamma^i)=-\log p_\theta\left(\mathbf{C}^i\right)=-\sum_{t=1}^{L_i} \log p_\theta\left(\mathbf{C}_t^i \mid \mathbf{C}_{<t}^i\right)
\end{equation}

The entropy (or average action) on the dataset $\Gamma'$ is:
\begin{equation}
    H_\theta(\Gamma') = -\sum_{\mathbf{C}^i \in \Gamma'} p_\theta\left(\mathbf{C}^i\right) \log p_\theta\left(\mathbf{C}^i\right)
\end{equation}

because we don't know $p(y_i)$ we use the ensemble average on all the trajectories in the dataset:
\begin{equation}
    H_\theta(\Gamma') = \frac{1}{N} \sum_{i=1}^N-\log p_\theta\left(\mathbf{C}^i\right)=\frac{-1}{N} \sum_{i=1}^N \sum_{t=1}^{L_i} \log p_\theta\left(\mathbf{C}_t^i \mid \mathbf{C}_{<t}^i\right)
\end{equation}

To model $p_\theta$ we utilize an \textbf{autoregressive Transformers neural network} \cite{vaswani2017attention} which is well-suited for capturing sequential dependencies and complex probability distributions. We minimize the \textbf{action} in the space of neural network parameters, which, in the context of machine learning, is equivalent to the \textbf{loss function}:

\begin{equation}
    \theta^* = \min_\theta H_\theta(\Gamma')
\end{equation}

in which $\theta^*$ are the learned parameters. For this modeling to be effective, we need to be able to sample from the cells distribution. To achieve this goal we take a simple approach of treating all the cells in the dataset as the vocabulary and then reduce the size of it using discretizing the cell space by Voroni diagrams.

\section{Data Availability}
We incorporated scRNA-seq data from
various datasets of cellular reprogramming 
\cite{schiebinger2019optimal}, spatiotempral mouse organogenesis development \cite{chen2022spatiotemporal} and mouse embryonic development \cite{qiu2024single} all available publically in the original papers. 

\section{Code Availability}
The training script for the model, tokenizer and the code for preprocessing data and inference are publicly available on the Github repository: \href{https://github.com/roholazandie/BioLeastAction}{https://github.com/roholazandie/BioLeastAction}

\section{Acknowledgement}
The work was supported in part by a grant from the National Institutes of Health R01 HL161069 awarded to ERE.


\section{Competing Interests Statement}
The authors declare no competing interests.


\end{document}